\documentstyle[preprint,prb,aps,epsf]{revtex}

\begin{document}
\tightenlines

\title{Electronic correlations, electron-phonon interaction, and 
isotope effect in high-$T_c$ cuprates}

\author{{A. Greco}$^a$ and {R. Zeyher}$^b$} 

\address{$^a$Departamento de F\'{\i}sica, Facultad de
Ciencias Exactas e Ingenier\'{\i}a and \\
IFIR(UNR-CONICET), Av. Pellegrini 250, 
2000-Rosario, Argentina\\
$^b$ Max-Planck-Institut\  f\"ur\
Festk\"orperforschung,\\ Heisenbergstr.1, 70569 Stuttgart, Germany}

\date{\today}

\vspace{5cm}

\maketitle

\begin{abstract} 
Using a large-$N$ expansion we present and solve the linearized
equation for the superconducting gap for a generalized $t-J$ model
which also contains phonons within a Holstein model. Keeping all terms
up to $O(1/N)$ the kernel of the gap equation consists of an
electron-phonon part with self-energies and vertex functions renormalized
by the interactions of the $t-J$ model, and a $t-J$ part unaffected by 
phonons. Considering first the electron-phonon part we find that
the leading superconducting instability always occurs in the 
s-wave channel with a $T_c$ which is lower compared to that of
non-interacting electrons, especially at large dopings.
The corresponding isotope coefficient $\beta$ is larger than
$1/2$ and increases with decreasing doping. Including also the $t-J$
part in the gap equation the leading $T_c$ has always d-wave symmetry
with phonons giving a positive contribution to $T_c$. $\beta$
is very small at small dopings and increases towards the classical 
value $1/2$ with increasing dopings similar as in many cuprates.
Considered as a function of the phonon frequency $\omega_0$
at a fixed coupling strength $\beta$ increases monotonously from
zero to about $1/2$ with increasing $\omega_0$.
\par
PACS numbers: 74.20.Mn, 74.25.Kc, 74.20-z
\end{abstract}
\newpage

\section{Introduction}

High-T$_c$ cuprates show strong electronic correlations but also a 
non-neglegible electron-phonon interaction. The latter causes, for
instance, the observed isotope effect in the transition temperature T$_c$,
especially, away from optimal doping \cite{Frank}. It is thus of interest to
calculate superconducting instabilities taking both the electron-electron
and the electron-phonon interaction into account. There are several
calculations dealing with the case of weak electronic correlations
\cite{Liechtenstein,Schuettler,Dahm,Bulut,Nunner,Pao}. In these calculations
the phonon-mediated part of the effective interaction 
is unaffected by electronic correlations. The
electronic part to the effective interaction is calculated 
using a Hubbard model and assuming $U$ to be small compared to the 
band width. It is repulsive and strongly peaked near the $M$-point in the 
Brillouin zone. As an alternative, Refs.\cite{Schuettler,Nunner}
use the spin-fluctuation term of Ref.\cite{Monthoux}. 
Since the phonon part is attractive throughout the Brillouin
zone it is plausible that the d-wave pairing due to the electronic part
will be weakened if the electrons couple strongly to phonons near the 
$M-$ point and only weakly affected if only phonons with small momentum
are involved \cite{Liechtenstein,Dahm,Bulut,Nunner,Jepsen}. 
Refs. \cite{Dahm,Pao}, for instance, find that the $T_c$
for d-wave superconductivity is always lowered by phonons and that the
isotope coefficient $\beta$
is negative for a constant electron-phonon coupling function. If only
phonons with small momenta play a role $\beta$ becomes positive\cite{Dahm}.
Ref. \cite{Nunner} even finds a change of sign in $\beta$ as a function
of the phonon frequency. 

The case of strong electronic correlations has quite different
features compared to the weak correlated case. The purely electronic
part of the effective interaction is rather a smooth function of momentum
without a sharp peak at the $M$-point due to spin fluctuations 
\cite{Zeyher1}. 
Furthermore, electronic correlations modify substantially the
phonon-mediated part of the effective interaction 
\cite{Kulic,Zeyher2,Zeyher3}.
It is the purpose of this communication to present results for this
case of strong electronic correlations. A suitable model for such
calculations is a generalized $t-J$ model which also includes phonons
within a Holstein model. Extending the spin degrees from 2 to $N$
systematic approximations for the effective interaction in terms of powers
in $1/N$ can be carried out \cite{Kulic,Zeyher2,Zeyher3,Ruckenstein}. 
In a first step we consider only the
phonon-mediated effective interaction which, unlike in the case of weak
correlations, is strongly modified by vertex corrections transforming the
original constant electron-phonon coupling of the Holstein model into a
strongly momentum- and frequency-dependent function. Explicit results
for $T_c$ and $\beta$ will be given for the leading symmetry channels
of the superconducting order parameter as a function of the doping. In a
second step these results are extended to the case where also the purely
electronic contribution to the effective interaction is taken into account.

\section{Linearized equation for the superconductivity gap}

Our Hamiltonian for the $t-J$ model plus phonons can be written as 
\begin{eqnarray}
H = \sum_{{ij} \atop {p=1...N}} {{t_{ij}} \over N} X_i^{p0} X_j^{0p}
+ \sum_{{ij} \atop {p,q=1...N}} {{J_{ij}} \over {4N}} X_i^{pq}
X_j^{qp} &-& \nonumber\\ 
\sum_{{ij} \atop {p,q=1...N}} {{J_{ij}} \over {4N}} X_i^{pp}  
X_j^{qq} 
+\sum_i \omega_0 (a^{\dagger}_i a_i+\frac{1}{2})&+& \nonumber\\
\sum_{i \atop {p=1...N}} \frac{g}{\sqrt{N}} 
[a^{\dagger}_i+a_i]({X_{i}}^{pp}-<{X_{i}}^{pp}>).
\end{eqnarray}

The first three terms correspond for $N=2$ to the usual $t-J$ model.
For $N=2$ $X$ is identical with Hubbard's projection operator $X_i^{pq}=
|{p \atop i}><{q \atop i}|$, where $|{p \atop i}>$ denotes for $p=0$ an
empty and for $p=1,2$ a singly occupied state with spin up and down,
respectively, at the site $i$. $t_{ij}$ and $J_{ij}$ are hopping
and Heisenberg interaction matrix elements between the sites $i,j$.
The fourth term in Eq.(1) represents one branch of dispersionless,
harmonic phonons with frequency $\omega_0$. The fifth term in Eq.(1)
describes a local coupling between the phonon and the change in the
electronic density at site $i$ with the coupling constant $g$.
$<X>$ denotes the expectation value of $X$. The extension from $N=2$
in Eq.(1) to a general $N$ has been discussed in detail in
Ref. \cite{Zeyher2}.
The label $p$ runs then not only over the two spin directions but also
over $N/2$ identical copies of the orbital. The symmetry group
of $H$ is the symplectic group $Sp(N/2)$ which allows to perform
$1/N$ expansions for physical observables. In Eq.(1) the electron-phonon
coupling is scaled as $1/\sqrt{N}$ whereas the free phonon part is
independent of $N$. As a result the leading contributions to 
superconductivity from the $t-J$ model alone and from the phonons
are both of order $O(1/N)$ which allows to treat them on an equal footing.

Instabilities towards superconductivity in the $t-J$ model have been studied
in the above framework in Refs. \cite{Zeyher1,Greco,Zeyher4}. 
The contributions to the anomalous
self-energy from the phonon-mediated effective interaction have been
derived in Refs. \cite{Kulic,Zeyher2} for the case $J_{ij}=0$. 
Generalizing the latter
treatment to a finite $J$ similar as in Ref. \cite{Zeyher3} the linearized 
equation for the superconducting gap $\Sigma_{an}$ for the entire
Hamiltonian Eq.(1) can be written as 

\begin{eqnarray}
{\Sigma}_{an}(k) = -{T \over {NN_c}}\sum_{k'}\Theta(k,k')
{1 \over {\omega_{n'}^2 +\epsilon^2({\bf k'})}}
{\Sigma}_{an}(k'). 
\end{eqnarray}
$N_c$ is the number of cells and $k$ the supervector $k=(\omega_n,{\bf k})$,
where $\omega_n$ denotes a fermionic Matsubara frequency and $\bf k$
the wave vector. $\epsilon({\bf k})$ is the one-particle
energy which is unchanged by the phonons in O(1) and thus given by 
Eq.(41) of Ref. \cite{Zeyher1}. The kernel $\Theta$ in Eq.(2) consists of
two parts
\begin{eqnarray}
\Theta(k,k') = \Theta^{t-J}(k,k')+\Theta^{e-p}(k,k').
\end{eqnarray}
The first contribution $\Theta^{t-J}$ in Eq.(3) comes from the 
$t-J$ model. Explicit expressions for it have been given in Eqs.(42)-(52)
in Ref. \cite{Zeyher1}. The second term $\Theta^{e-p}$ in Eq.(3) is due to 
phonon-mediated interactions and is given by 
\begin{eqnarray}
\Theta^{e-p}(k,k') &=&  - \frac{2g^2\omega_0}{(\omega_{n}-
\omega_{n'})^2+{\omega_0}^2} \nonumber\\
&\cdot&{\gamma_{c}(k',k-k')}{\gamma_{c}(k,k'-k)}.
\end{eqnarray}
$\gamma_c$ is the charge vertex for which an explicit expression has
been derived in Ref. \cite{Zeyher3}. For uncorrelated electrons there are no
vertex corrections, i.e., $\gamma_c=-1$. Since $g$
and $\omega_0$ are assumed to be independent of $\bf k$ $\Theta^{e-p}$
is also independent of $\bf k$ and only s-wave superconductivity
is possible. When correlations are present the kernel $\Theta^{e-p}$ 
depends on $\bf k$ and $\omega_n$ through the charge vertex $\gamma_c$
and general symmetries for the order parameter may become possible.

\section{Calculation of $T_c$ and $\beta$ from the phonon-mediated part}

Taking a square lattice with point group $C_{4v}$ $\Theta(k,k')$ and
$\epsilon({\bf k})$ in Eq.(2) are invariant under $C_{4v}$ which
means that $\Sigma_{an}(k)$ can be classified according to the five
irreducible representations $\Gamma_i$ of $C_{4v}$. s-wave symmetry 
corresponds to $\Gamma_1$, d-wave symmetry to $\Gamma_3$, etc.
In the weak-coupling case $\Theta(k,k')$ can be approximated
by its static limit $\Theta({\bf k},{\bf k'})$. Putting all momenta
right onto the Fermi line the sum over 
${\bf k'}$ in Eq.(2) can be transformed into a line integral along the 
Fermi line.
Assuming a certain irreducible representation $\Gamma_i$ for the 
order parameter,
the line integral can be restricted to the irreducible Brillouin zone (IBZ)
introducing a symmetry-projected kernel $\Theta_i({\bf k},{\bf k'})$
with ${\bf k},{\bf k'} \in$ IBZ. Finally, the line integral was discretized
by a set of points $[{\bf k}^F_\alpha]$ along the Fermi line in the IBZ
with line elements $[s({\bf k}^F_\alpha)]$. Denoting the smallest
eigenvalue of the symmetric matrix
\begin{equation}
{1 \over {4 \pi^2}} \sqrt{ {s({\bf k}_\alpha^F)s({\bf k}^F_\beta)} \over
{|\nabla \epsilon({\bf k}^F_\alpha)| \cdot |\nabla \epsilon({\bf k}^F_\beta)|}}
\Theta_i({\bf k}^F_\alpha,{\bf k}^F_\beta) 
\end{equation}
by $\lambda_i$ Eq.(2) yields for $\lambda_i <0$, $N=2$, and in the
weak-coupling case the BCS-formula
\begin{equation}
T_{ci} = 1.13 \omega_0 e^{1/ \lambda_i} . 
\end{equation}
As usual we took $\omega_0$ as a suitable cutoff. If $\lambda_i >0$ we have, 
of course, $T_{ci}=0$. According to Eq.(6)
the absolute value of $\lambda_i$ characterizes the strength of the 
effective interaction in the symmetry channel $i$. The overall strength
of the electron-phonon coupling is conventionally expressed in terms
of the dimensionless coupling $\lambda$ defined by
\begin{eqnarray}
\lambda=\frac{g^2}{8 \omega_0}.
\end{eqnarray}
In Eq.(7) we have introduced a factor $1/2$
to account for the prefactor $1/2$ in Eq.(2) after setting there $N=2$.
We also used the average density of states $1/(8t)$ for the density
of states factor in $\lambda$ and put $t$ equal to 1. For the range of
dopings we will be interested in, i.e., $0.15 < \delta < 0.8$,
the density of states varies only little with doping so that the 
definition Eq.(7) of $\lambda$ is appropriate for all dopings.

\vspace{-0.5cm}
\begin{figure}
      \epsfysize=75mm
      \centerline{\epsffile{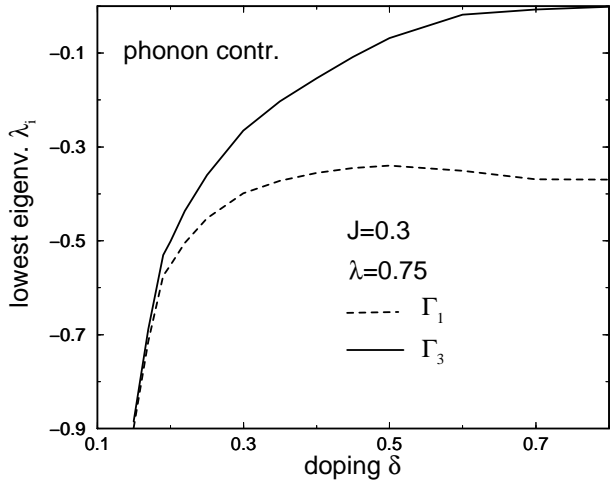}}
\caption{
Phonon contribution to the lowest eigenvalues $\lambda_i$
of the static kernel $\Theta$ for the representations $\Gamma_1$ and $\Gamma_3$
of 
$C_{4v}$ as a function of the doping $\delta$.}
\label{fig:fig1}
\end{figure}

Fig. 1 shows the eigenvalues $\lambda_i$ for the two leading symmetries
$i=1,3$ using the parameter value $J=0.3$ and $t$
as the energy unit. $\delta$ is the doping measured from half-filling.
For $\lambda$ we have chosen the value $0.75$ which is a typical value 
obtained in LDA calculations \cite{Andersen,Krakauer}. The lowest eigenvalue
occurs always in s-wave symmetry, i.e., for $i=1$. It depends only
weakly on doping for $0.3 < \delta < 0.8$. In this region the order
parameter also varies only slowly along the Fermi line, describing thus
rather isotropic s-wave superconductivity. Below $\delta = 0.3$ $\lambda_1$ 
rapidly
decreases with decreasing doping wich is caused by a soft mode which
freezes into an incommensurate bond-order wave at  
$\delta_{BO} \sim 0.14$ \cite{Zeyher1}. As a result the s-wave order 
parameter becomes less and less isotropic with decreasing $\delta$
changing, for instance, for $\delta =0.2$ by about a factor 3 along
the Fermi line but it does not pass through zero. The d-wave coupling constant 
$\lambda_3$ would be zero in the uncorrelated case. Decreasing $\delta$
from large values correlation effects increase and the static vertex
function $\gamma({\bf k},{\bf k-k'})$ develops more and more a forward 
scattering peak in the transferred momentum $\bf k-k'$ 
\cite{Kulic,Zeyher2}. In the extreme case where $\gamma$ is proportional to
$\delta({\bf k-k'})$ $\lambda_1$ and $\lambda_3$ would become degenerate.
Fig. 1 shows that this degeneracy is nearly reached at low dopings.
The strong localization of the effective interaction is caused
by the forward scattering peak in $\gamma$ and, in addition, by the
incipient instability at $\delta_{BO}$ which is also very localized in
${\bf k}$-space. 

The neglect of retardation effects which leads to the BCS-formula Eq.(6)
is doubtful for several reasons. First, $\lambda = 0.75$ no longer
corresponds really to a weak-coupling case. Secondly, $T_c$ is
no longer small compared to the frequency of the soft mode near
$\delta_{BO}$ which means that the frequency dependence of $\Theta$
cannot be neglected for these dopings. Thirdly, and most severely,
the formation of the forward scattering peak in the
static vertex function $\gamma$ is accompagnied by a strong
frequency dependence which should be taken into account on the same footing
as its momentum dependence. We thus have solved the gap equation
Eq.(2) assuming only that the momenta can be put right onto the
Fermi line but keeping the full frequency and momentum
dependence along the Fermi line. 

\vspace{-0.5cm}
\begin{figure}
      \epsfysize=75mm
      \centerline{\epsffile{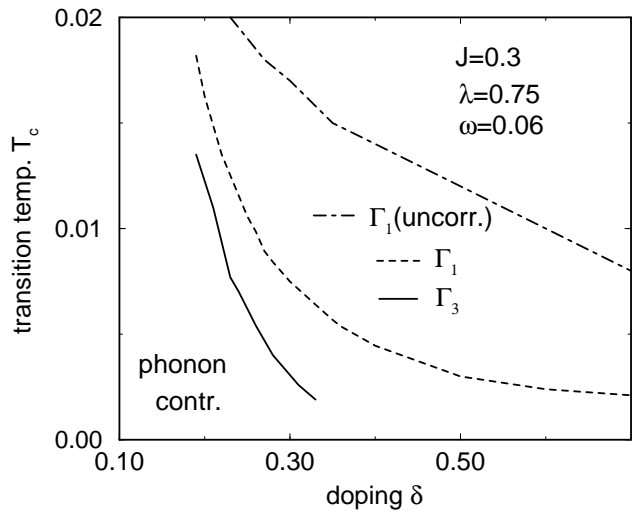}}
\caption{
Transition temperature $T_c$ in units of $t$ as a function
of the doping $\delta$ for the uncorrelated (dash-dotted line,
$\Gamma_1$ symmetry) and correlated (dashed and solid lines for $\Gamma_1$ 
and $\Gamma_3$ symmetries, resp.) cases, using only the phonon 
contribution.}
\label{fig:fig2}
\end{figure}

Fig. 2 shows the obtained results for
$T_c$ for a s-wave (dashed line) and a d-wave (solid line) order
parameter, together with the $T_c$ curve for the uncorrelated case 
(dash-dotted line). The first observation is that correlation effects
always suppress s-wave superconductivity. The suppression is
large for $\delta > 0.25$ and becomes less pronounced at smaller
dopings in agreement with previous results based on the static 
approximation \cite{Zeyher2}. The order parameter associated with
the dashed line in Fig. 2 varies only slowly along the Fermi line,
i.e., we have a usual s-wave order parameter without nodes. In
a Gutzwiller description this order parameter would be exactly zero
if retardation effects can be neglected and $N=2$ is taken: 
Integrating out the phonons the effective interaction becomes in the 
static limit proportional to the double occupancy operator.
Any matrix element formed with Gutzwiller wave functions thus would 
be zero. In contrast to that the enforcement of our constraint at large 
$N$'s, namely, that only $N/2$ out of the total $N$ states at a given
site can be occupied at the same time, gives rise to another effect,
which is absent in the Gutzwiller treatment: the effective interaction
becomes more and more long-ranged with decreasing doping.
This makes isotropic s-wave superconductivity possible even in the presence
of a $N=2$ constraint. Fig. 2 nevertheless shows that the reduction of
$T_c$ due to the constraint is substantial except at very small dopings
where the effective interaction becomes extremely long-ranged
and the suppression by the constraint is small. 

The kernel $\Theta^{e-p}$ is in the absence of correlations and in the
static limit negative for all arguments $\bf k$ and $\bf k'$. This implies
that the s-wave symmetry has always the highest $T_c$. According
to our numerical studies the same is
also true in the correlated case which explains 
why the solid line in Fig. 2, describing d-wave
superconductivity, is always below the dashed line. As indicated previously
the very existence of finite values for $T_c$ with d-wave symmetry
is in our model due to electronic correlation effects. Fig. 1 showed that
the momentum dependence of $\gamma_c$ causes in the static limit a finite
coupling strength in the d-wave channel. The solid line in Fig. 2
proves that a finite $T_c$ in the d-wave channel results from this even
if the strong frequency dependence of $\gamma_c$ is also taken into account.

\vspace{-0.5cm}
\begin{figure}
      \epsfysize=75mm
      \centerline{\epsffile{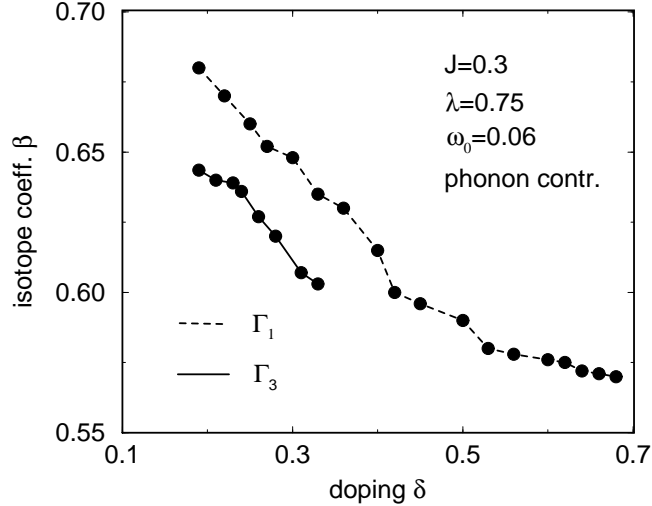}}
\caption{
Isotope coefficient $\beta$ as a function of the doping 
$\delta$ considering only the phonon contribution for $\Gamma_1$ (dashed 
line) and $\Gamma_3$ (solid line) symmetries. The filled circles are 
calculated values.}
\label{fig:fig3}
\end{figure}

Fig. 3 shows results for the isotope coefficient $\beta$ in the 
case of s-wave (dashed line) and d-wave (solid line) superconductivty.
We assumed hereby that $\omega_0$ is inversely and $g$ directly 
proportional to the 
the square root of the mass rendering $\lambda$ independent of the mass.
Neglecting the frequency dependence of $\gamma_c$ there is only
one energy scale involved in the gap equation which yields
$\beta = 1/2$ independent of doping, $\lambda$, etc.

In contrast to that
the curves in Fig. 3
show that $\beta$ increases monotonically with decreasing doping
and is always larger than 1/2. The deviaton of $\beta$ from the value 1/2
is due to the frequency dependence of $\gamma_c$ which increases with
decreasing doping \cite{Zeyher1} and acts as a second energy scale. The
non-adiabatic effects produced by electronic correlations
thus always increase $\beta$ for our range of parameter values.
This should be contrasted to the case where vertex corrections
due to the electron-phonon interaction has been studied 
\cite{Grimaldi,Miller}, and where $\beta$
may be larger or smaller than $1/2$. 
We can conclude from our results that models 
based on phonon-induced effective interactions in the
presence of correlations seem not to be very attractive models for 
high-T$_c$ superconductivity because
a) the largest transition temperatures are found for s-wave symmetry
for all dopings and 
b) the isotope coefficient is larger than $0.5$, especially at
low dopings. Experimentally, high-$T_c$ cuprates exhibit
d-wave superconductivity and, at least near optimal dopings, small isotope 
effects.

\section{Calculation of $T_c$ and $\beta$ from the full effective
interaction}

Superconducting instabilites of the pure $t-J$ model have been studied in 
\cite{Zeyher1,Greco,Zeyher4}. The smallest eigenvalue of $\Theta^{t-J}$
and, correspondingly, the largest $T_{c}$ always occur in the d-wave 
symmetry channel. On the other hand,
the phonon-induced effective interaction described by $\Theta^{e-p}$
favors in the $t-J$ model mainly s-wave superconductivity. In this section,
we will consider the case where both effective interactions are present
and study the symmetry of the resulting superconducting state, its
transition temperature and isotope coefficient.

\vspace{-0.5cm}
\begin{figure}
      \epsfysize=75mm
      \centerline{\epsffile{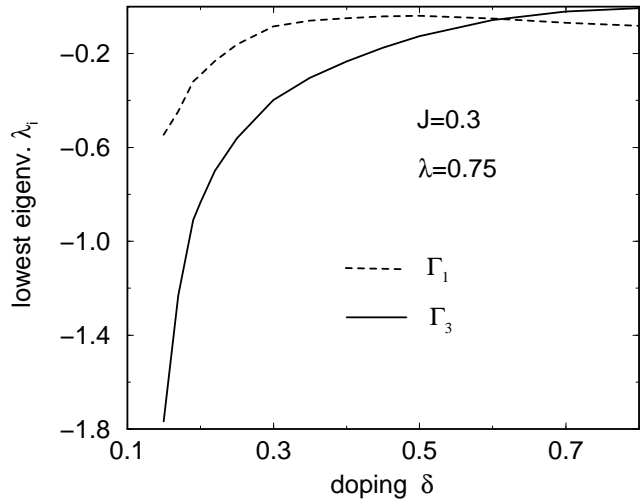}}
\caption{
Lowest eigenvalues $\lambda_i$ of the total static kernel $\Theta$ 
for the representations $\Gamma_1$ and $\Gamma_3$ of $C_{4v}$ as a function 
of the doping $\delta$.}
\label{fig:fig4}
\end{figure}

Fig. 4 shows the two lowest eigenvalues of the matrix Eq.(5)
which occur in the s- and d-wave symmetry channel. For $\Theta$
both terms in Eq.(3) have been included. For $\delta <  0.6$
the lowest eigenvalue has d-wave symmetry and decreases steadily
and strongly with decreasing doping diverging finally at $\delta_{BO}$.    
Comparing the solid lines in Figs. 1 and 4 of this work and Fig. 2 in 
\cite{Zeyher1}
one sees that the lowest eigenvalue is roughly additive in the
$t-J$ and the phonon contributions though, of course, $\Theta^{t-J}$
and $\Theta^{e-p}$ do not commute with each other. Note also the 
considerable contribution of the phonon part to $\lambda_3$ though the
overall dimensionless coupling constant has the rather moderate
value $\lambda = 0.75$. The eigenvalue $\lambda_1$ (dashed line in
Fig. 4) is rather constant for $\delta > 0.3$ and decreases
much less towards lower dopings than $\lambda_3$. There is a crossover
from d-wave to s-wave symmetry in the lowest eigenvalue at $\delta \sim 0.6$
suggesting that the s-wave order parameter is more stable than the
d-wave order parameter at high dopings. Such a crossover is expected
for general reasons: For large dopings correlation effects play
a minor role. As a result the d-wave part in the phonon-induced
effective interaction as well as the whole $t-J$ part becomes small
whereas the s-wave part of the Holstein model 
will dominate. A comparison of the solid lines in 
Figs. 1 and 4 nevertheless shows that the latter is very effectively 
suppressed by the $t-J$  part at practically all considered dopings.
For instance, $|\lambda_1|$ in Fig. 1 is for $\delta > 0.25$
about a factor 4 larger than in Fig. 4. 

\vspace{-0.5cm}
\begin{figure}
      \epsfysize=75mm
      \centerline{\epsffile{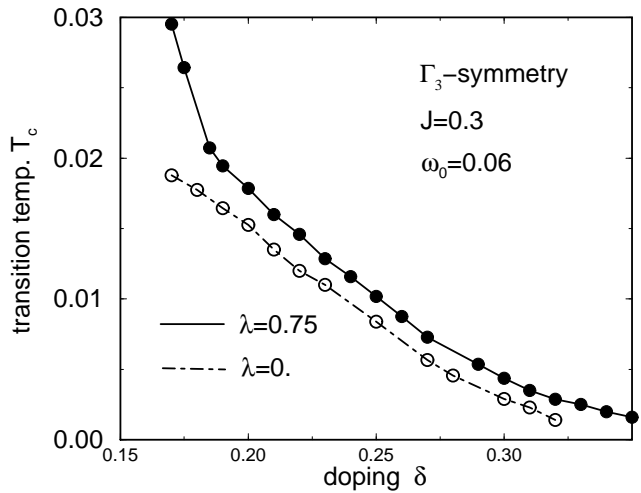}}
\caption{
Transition temperature $T_c$ in units of $t$ as a function
of the doping $\delta$ for $\Gamma_3$ symmetry for $\lambda=0$ 
(dash-dotted  
line) and $\lambda =0.75$ (solid line).}
\label{fig:fig5}
\end{figure}

The solution of the gap equation Eq.(2) is not straightforward because
$\Theta$ contains both instantaneous and retarded contributions and
because the various retarded contributions have different effective
cutoff energies. We developed in \cite{Zeyher1} a method to solve Eq.(2)
directly avoiding the use of pseudopotentials. The only simplification
is that we put the momenta in retarded (but not in instantaneous)
contributions to the Fermi line. This approximation
has been checked numerically and found to be very well satisfied in
our case. Using the fact that the instantaneous kernel consists only
of a few separable contributions Eq.(2) can for a given Matsubara
frequency reduced to a linear matrix problem of order 1 and 3 for d- and 
s-wave symmetry, respectively. 

In Fig. 5 we have plotted the calculated values for $T_c$ for d-wave
symmetry $\Gamma_3$ as a function of the doping $\delta$. The dashed-dotted
line corresponds to $\lambda=0$, the solid line to $\lambda=0.75$.
The Figure shows that phonons always increase the $T_c$ for 
d-wave superconductivity. The increase is especially large at low
dopings. This is quite in contrast to calculations in the weak-coupling
case \cite{Dahm,Nunner} where the $T_c$ of d-wave superconductivity
is lowered by phonons. The physical picture in the latter case
is that the effective interaction of the electronic part
is repulsive and strongly peaked near the $M$-point whereas the
phononic d-wave part is attractive and diminuishes especially at
the $M$-point the repulsion causing a lowering of $T_c$. In our case
the dependence of the static effective interaction of the pure $t-J$ model
on the transferred momentum can be inferred from Figs.1a) and 1b) in 
Ref. \cite{Zeyher1}. For small dopings the phonon part
is restricted to small momenta which means that it would contribute
only near the $X$-point in Fig. 1b) of Ref. \cite{Zeyher1}. As a result
the total effective interaction would become even more attractive
around the $X$-point and elsewhere not be changed. This clearly would
enhance the d-wave part of the effective interaction which is also in
agreement with Fig. 4. From Figs. 1, 4 and 5 it is evident that the
large lowering of the eigenvalue $\lambda_3$ due to phonons
corresponds only to a rather moderate increase in $T_c$. This means
that the use of a BCS-formula with a fixed effective cutoff would grossly 
overestimate the increase in $T_c$ for d-wave superconductivity due to 
phonons. What has been overlooked in such an approach is that we deal with
at least three energy scales, namely $t$, $J$, and $\omega_0$. For 
instance, the
instantaneous contribution of the $t-J$ part is characterized by the energy
scale $t$ whereas the phononic one by $\omega_0$. Since $t >> \omega_0$
it is evident that the phononic part in $\lambda_3$ will contribute 
to $T_c$ much less than the $t-J$ part. 

We were unable to find any finite transition temperatures $T_{ci}$
for $i \neq 3$, i.e., for symmetries different from the d-wave symmetry..
Taking the accuracy of our calculation into account this means that
$T_{ci} < 0.002$ for $i \neq 3$. As shown in Fig. 2 the
phonon-induced effective interactions leads to a considerable
$T_c$ for s-wave superconductivity. The $t-J$ part
to the effective interaction, however, is very repulsive in the
s-wave channel prohibiting
s-wave superconductivity. We cannot exclude that the crossover 
at $\delta \sim 0.6$ in Fig. 4 in the lowest eigenvalue from d-wave 
to s-wave could stabilize a s-wave order parameter at large dopings.
Another possibility is that other symmetries than d- or s-wave become stable 
at large dopings in view of the approximate degeneracy of 
the eigenvalues of practically all symmetries in the pure $t-J$ model
\cite{Zeyher1}.
In any case, the corresponding $T_c$'s  would 
be smaller than $\sim 0.002$ for all dopings and thus be rather
irrelevant.

The behavior of the isotope coefficient $\beta$ as a function of
doping is another interesting test for any theory of high-$T_c$
superconductivity. Fig. 6 shows the calculated $\beta$ for
$\lambda = 0.75$, $J=0.3$, $\omega_0=0.06$ in the case of $\Gamma_3$ symmetry.
It is always positive, starting from small values at small dopings.
With increasing doping it increases monotonously approaching values near 
$1/2$ at large dopings. Our calculated curve shows the typical behavior
of the experimetal data, namely, a small value for $\beta$ at optimal 
doping and substantial values far away from optimal doping \cite{Frank}.
The curve in Fig. 6 is the result of several 

\vspace{-0.5cm}
\begin{figure}
      \epsfysize=75mm
      \centerline{\epsffile{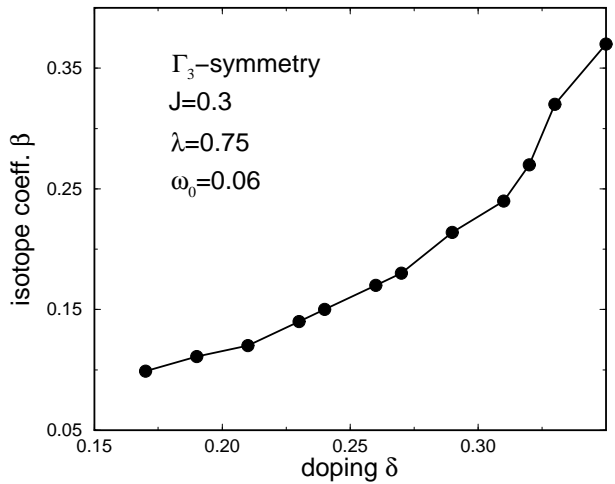}}
\caption{
Isotope coefficient $\beta$ as a function of the doping 
$\delta$
for d-wave-like $\Gamma_3$ symmetry. The filled circles are
calculated values.}
\label{fig:fig6}
\end{figure}
competing effects. We saw in Fig. 3 that correlation effects in the 
phonon-induced part
cause a doping dependence of $\beta$ which is opposite to that in
Fig. 6, namely, a monotonously increasing $\beta$ with decreasing doping
starting at the classical value of $1/2$ at large dopings. Including also
the $t-J$ part the large values of $\beta$ at low dopings are nearly
completely and at larger dopings partially quenched. Part of this effect may 
be understood from Fig. 5. It shows that the relative importance
of the phonon contribution increases steadily with increasing $\delta$
except at very small dopings. In a simple picture one thus may 
argue that at small dopings
$T_c$ is mainly due to the $t-J$ part. This part decays faster
than the phononic part with doping so that $T_c$ at larger dopings     
is mainly due to the phonons. A more realistic interpretation of Fig. 6
should, however, also take into account the complex competition
between the three energy scales $t$, $J$, and $\omega_0$, pair
breaking effects, etc.

\begin{figure}
      \epsfysize=75mm
      \centerline{\epsffile{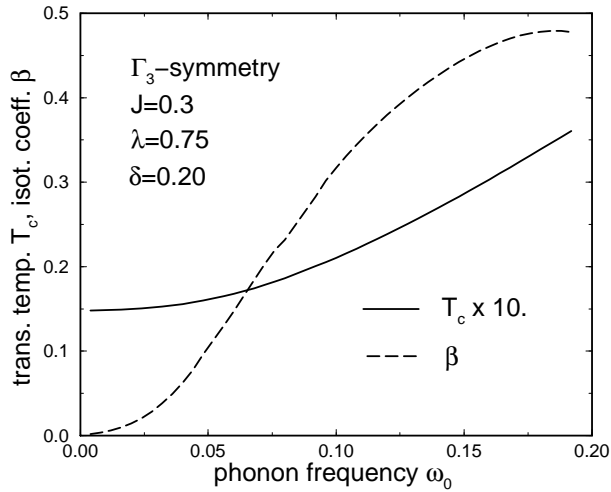}}
\caption{
Transition temperature $T_c$, multiplied by 10, and 
isotope coefficient $\beta$ as a function of the phonon frequency
$\omega_0$ for d-wave-like $\Gamma_3$ symmetry.}
\label{fig:fig7}
\end{figure}

In order to get more insight into the behavior of $\beta$ we
have calculated $T_c$ and $\beta$ in the $\Gamma_3$ symmetry channel
for the fixed doping value $\delta = 0.20$ as a function of the phonon
frequency $\omega_0$. Fig. \ref{fig:fig7} shows the result for $J = 0.3$
and $\lambda = 0.75$. The solid line in Fig. \ref{fig:fig7} represents the 
value
for $T_c$, multiplied with 10. Since we keep $\lambda$ fixed the
coupling constant $g$ approaches zero in the adiabatic limit 
$\omega_0 \rightarrow 0$ which means that $T_c$ in this limit is
solely due to the $t-J$ part. With increasing phonon frequency $T_c$

\vspace{-0.5cm}
increases monotonously, passing over from an initial sublinear
to a linear behavior. The solid line shows that phonons always increase
$T_c$ for d-wave superconductivity. The absolute values in the
Figure are quite remarkable:  Though the employed coupling constant 
for $\lambda$ is rather moderate and though the phonon contribution to 
$T_c$ in the d-wave channel is entirely due to correlation effects
$T_c$ increases by nearly a factor 3 for $\omega_0 =0.2$ corresponding to
about the largest phonons in the cuprates. 
The dashed line in Fig. \ref{fig:fig7} represents the
dependence of $\beta$ on $\omega_0$. For small $\omega_0$ $\beta$
is practically zero. $\beta$ increases monotonously with increasing
$\omega_0$ and tends to the classical value $1/2$ at large
phonon frequencies. The dashed line in Fig. \ref{fig:fig7} may
be understood in physical terms roughly in the following way. In our
model the static  effective coupling is independent of the phonon mass,
i.e., of $\omega_0$. $\beta$ is thus determined by an effective energy cutoff.
For small $\omega_0$'s this cutoff is mainly given by electronic
parameters in the $t-J$ part leading to a small value for $\beta$.
At large $\omega_0$'s the phonon contribution to $T_c$, which is only due
to electronic correlations, determines  mainly the total effective energy 
cutoff causing a value of $\beta$ near $1/2$.

In conclusion, we have treated the electron-electron and the 
electron-phonon interactions in a generalized $t-J$ model by means of a
systematic $1/N$ expansion and have solved the resulting 
linearized equation for
the superconducting gap by reliable numerical methods. We found that
electronic correlations affect phonon-induced superconductivity in
several ways: Instabilities towards d-wave or other symmetries different
from s-wave become possible. The corresponding transition temperatures
are, however, always smaller than that of s-wave superconductivity.
Moreover, the $T_c$ for s-wave superconductivity is at larger
dopings heavily and at small dopings somewhat suppressed
by correlations. The isotope coefficient $\beta$ is $1/2$ at large
dopings but increases with increasing correlations, i.e., decreasing
dopings, both in the s- and d-wave channel. Including also the $t-J$ part
in the effective interaction we found 
that within our numerical accurracy only d-wave superconductivity is
stable for dopings $0.15 < \delta < 0.8$ and that the phononic part
always increases $T_c$ for the above dopings. The $t-J$ part in the
effective interaction changes the dependence of $\beta$ on doping:
$\beta$ assumes now small values at low dopings and increases monotonously
with doping towards the classical value $1/2$ at large dopings. 
Keeping $\lambda$ fixed and varying the phonon mass, i.e., $\omega_0$
we find that $T_c$ and $\beta$ increase monotonously with $\omega_0$
and that $\beta$ varies between zero at small and roughly $1/2$
at large phonon frequencies within the interval $0 < \omega_0 < 0.2$. 

Our findings differ in many aspects from 
the corresponding results based on weak-coupling 
calculations \cite{Schuettler,Dahm,Bulut,Nunner,Pao}. In these calculations
$T_c$ for d-wave superconductivity induced either by a $U$ or a 
spin-fluctuation term is always suppressed by phonons. For instance,
in the fluctuation-exchange approximation $T_c$ drops to zero if the
phonon-mediated on-site attraction $U_p$ becomes comparable to the
Hubbard term $U$\cite{Pao}. 
We treated the opposite case where $U >> U_p$ and found
a different behavior, namely, that phonons enhance the $T_c$ for d-wave
superconductivity. Our different result is mainly caused by corrections to the
bare electron-phonon vertex due to the strong electron-electron interaction.
This vertex develops for not too large dopings a forward scattering peak
so that only phonon with small momenta can couple to the electrons.
As a result the electron-phonon and the $t-J$ contributions to the gap
decouple in $\bf k$ space. Our calculations show that the two
contributions no longer cancel each other to a large extent but,
on the contrary, enhance each other. For a momentum-independent bare
electron-phonon coupling weak-coupling calculations yield a small, often
negative value for the isotope coefficient $\beta$ which is rather 
independent of doping. In our case the strong momentum-dependence
of the effective electron-phonon coupling, induced by electronic
correlations, causes a strong dependence of $\beta$ on doping:
being always positive, $\beta$ is small at optimal doping and assumes values
of roughly $1/2$ at large dopings in agreement with 
the basic features of the experimental data in the cuprates. 

The presented results are accurate at large $N$'s because we have taken
only the leading terms of a $1/N$ expansion. Phonon renormalizations and
vertex corrections due to the electron-phonon interaction are of $O(1/N)$
and thus have been omitted. On the other hand it is known that
in the physical case $N=2$ anharmonic effects in the atomic potentials
and the formation of polarons occur if $\lambda$ is about 1 or larger
\cite{Pao1}. This suggests that keeping only the leading order of the
$1/N$ expansion cannot describe adequately the case $N=2$ at large
coupling strengths or if the Migdal ratio $\omega_0/t$ is no longer small.
Correspondingly, we have shown numerical results for rather moderate
values for $\lambda$ and small Migdal ratios.

{\bf Acknowledgement}: The authors are grateful to Secyt and the International
Bureau of the Federal Ministry for Education, Science, Research and 
Technology of Germany for financial support (Scientific-technological 
cooperation between Argentina and Germany, Project No. ARG AD 3P).
The first and second author thank the MPI-FKF, Stuttgart, Germany, and
the Departamento de F\'{\i}sica, Fac. Cs. Ex. e Ingenier\'{\i}a, U.N. Rosario,
Argentina, respectively, for hospitality. The authors also thank P. Horsch
for a critical reading of the manuscript. 


\vspace{2cm}


\begin{thebibliography}{99}

\bibitem{Frank} C. Frank, in Physical Properties 
of High-Temperature Superconductors I, ed. by D. M. Ginsberg
(World Scientific, Singapure, 1993), p.189
 
\bibitem{Liechtenstein} A.I. Lichtenstein and M.L. Kuli$\acute{c}$,
Physica {\bf C245}, 186 (1995)

\bibitem{Schuettler} H.-B. Sch\"uttler and C.-H. Pao, Phys. Rev. Lett.
{\bf 75}, 4504 (1995)

\bibitem{Dahm} T. Dahm, D. Manske, D. Fay, and L. Tewordt, Phys. Rev. B
{\bf 54}, 12006 (1996)

\bibitem{Bulut} N. Bulut and D.J. Scalapino, Phys. Rev. B {\bf 54},
14971 (1996)
 
\bibitem{Nunner} T.S. Nunner, J. Schmalian, and K.H. Bennemann,
cond-mat/9804088

\bibitem{Pao} C.-H. Pao and H.-B. Sch\"uttler, Phys. Rev. B{\bf 57},
5051 (1998)

\bibitem{Monthoux} P. Monthoux and D. Pines, Phys. Rev. B{\bf 49},
4261 (1994)

\bibitem{Jepsen} S. Savrasov and O.K. Andersen, Phys. Rev. Lett. {\bf 77},
4430 (1996)

\bibitem{Zeyher1} R. Zeyher and A. Greco, Eur. Phys. J. B {\bf 6}, 473 (1998)

\bibitem{Kulic} M.L. Kuli\'c and R. Zeyher, Phys.Rev. B {\bf 49}, 
4395 (1994)

\bibitem{Zeyher2} R. Zeyher and M.L. Kuli\'c, Phys.Rev. B {\bf 53}, 
2850 (1996)

\bibitem{Zeyher3} R. Zeyher and M.L. Kuli\'c, Phys. Rev. B {\bf 54}, 8985
(1996)

\bibitem{Ruckenstein} A.E. Ruckenstein and S. Schmidt-Rink, Phys. Rev. B
{\bf 38}, 7188 9(1988)

\bibitem{Greco} A. Greco and R. Zeyher, Europhys. Lett. {\bf 35},115 (1996)

\bibitem{Zeyher4} R. Zeyher and A. Greco, Z. Physik B {\bf 104}, 737 (1997)

\bibitem{Andersen} O. K. Andersen, S.Y. Savrasov, O. Jepsen, and A.I.
Liechtenstein, J. of Low Temp. Physics {\bf 105}, 285 (1996)

\bibitem{Krakauer} {H. Krakauer, W.E. Pickett, and R.E. Cohen, Phys. Rev.\\
B {\bf 47}, 1002 (1993)}

\bibitem{Grimaldi} C. Grimaldi, L. Pietronero, and S. Str\"assler, Phys.
Rev. B {\bf 52}, 10530 (1995)

\bibitem{Miller} P. Miller, J.K. Freericks, and E.J. Nicol, 
cond-mat/9805254

\bibitem{Pao1} C.-H. Pao and H.-B. Sch\"uttler, Phys. Rev. Lett. {\bf 69},
1600 (1992)

\end{thebibliography}
\end{document}